\documentclass[aps,prl,twocolumn,superscriptaddress,english,floatfix]{revtex4-2}
\usepackage{graphicx}
\usepackage{float}
\usepackage{physics}
\usepackage{cancel}
\usepackage{overpic}
\usepackage{enumerate}
\usepackage{dsfont}
\usepackage{bbm}
\usepackage{mathrsfs}

\usepackage[colorlinks,citecolor=blue,linkcolor=blue,urlcolor=blue]{hyperref}
\usepackage{textcomp}
\usepackage{amsmath}
\usepackage{amssymb}
\usepackage{soul}
\usepackage[normalem]{ulem}
\usepackage{lipsum}
\usepackage{comment}
\usepackage{mathrsfs}

\usepackage[english]{babel}
\usepackage{bm}
\usepackage{orcidlink}

\graphicspath{{./}{./figs/}}

\let\originalleft\left
\let\originalright\right
\renewcommand{\left}{\mathopen{}\mathclose\bgroup\originalleft}
\renewcommand{\right}{\aftergroup\egroup\originalright}

\newcommand{\vect}[1]{\boldsymbol{#1}}
\renewcommand{\vec}[1]{\vect{#1}}

\def\rmi{{\rm {i}}}
\renewcommand{\d}{{\rm {d}}}

\newcommand{\idhat}{\hat{\mathds{1}}}
\newcommand{\rhohat}{\hat{\rho}}
\newcommand{\Omegavec}{\vec{\Omega}}
\newcommand{\thetavec}{{\vec{\theta}}}
\newcommand{\sigmahat}{\hat{\sigma}}

\newcommand{\Ahat}{\hat{A}}
\newcommand{\Fvec}{\vec{F}}
\newcommand{\Gcal}{\mathcal{G}}
\newcommand{\Ical}{\mathcal{I}}

\newcommand{\Hhat}{\hat{H}}
\newcommand{\lcal}{\mathcal{L}}

\newcommand{\mvec}{\vec{m}}
\newcommand{\ncal}{\mathcal{N}}
\newcommand{\nvec}{{\vec{n}}}

\newcommand{\Shat}{\hat{S}}

\newcommand{\bea}{\begin{equation}\begin{aligned}}
		\newcommand{\eea}{\end{aligned}\end{equation}}
\newcommand{\be}{\begin{equation}}
	\newcommand{\ee}{\end{equation}}

\begin{document}

\title{Variational Dynamics of Open Quantum Spin Systems in Phase Space}

\author{Jacopo Tosca~\orcidlink{0009-0002-5165-7937}}
\thanks{These authors contributed equally.}
\affiliation{Universit\'{e} Paris Cit\'e, CNRS, Mat\'{e}riaux et Ph\'{e}nom\`{e}nes Quantiques, 75013 Paris, France}

\author{Zejian Li~\orcidlink{0000-0002-5652-7034}}
\thanks{These authors contributed equally.}
\affiliation{Universit\'{e} Paris Cit\'e, CNRS, Mat\'{e}riaux et Ph\'{e}nom\`{e}nes Quantiques, 75013 Paris, France}

\author{Francesco Carnazza \orcidlink{0000-0002-1458-8701}} 
\affiliation{Universit\'{e} Paris Cit\'e, CNRS, Mat\'{e}riaux et Ph\'{e}nom\`{e}nes Quantiques, 75013 Paris, France}

\author{Cristiano Ciuti \orcidlink{0000-0002-1134-7013}}
\affiliation{Universit\'{e} Paris Cit\'e, CNRS, Mat\'{e}riaux et Ph\'{e}nom\`{e}nes Quantiques, 75013 Paris, France}
\begin{abstract}
We introduce a variational method for simulating the dynamics of interacting open quantum spin systems. The method is based on the spin phase-space representation and variationally targets the Husimi-$Q$ function with an ansatz based on a multi-dimensional mixture of spin-coherent states. Crucially, the mixture coefficients are allowed to take negative values, enabling the faithful capture of quantum correlations beyond semiclassical descriptions. The resulting equations of motion are derived from the Dirac-Frenkel variational principle and can be evaluated efficiently without resorting to Monte Carlo sampling by exploiting the analytical structure of the ansatz. As a first application, we demonstrate that this approach accurately captures both the full quantum dynamics and the non-equilibrium steady states of the transverse-field quantum Ising model, in excellent agreement with exact diagonalization. Furthermore, we show that the method scales efficiently to large two-dimensional lattices, a regime that remains challenging for other techniques.
\end{abstract}

\maketitle

{\it Introduction.---}
The realization of scalable quantum technologies is one of the main challenges of modern physics. A wide range of platforms, including quantum annealers \cite{Kadowaki1998, Farhi2001, Johnson2011}, Ising machines \cite{Kim2010, Wang2013, Goto2019, tosca2025-EmergentEquilibrium}, and universal quantum computers \cite{Preskill2018, Bluvstein2023, Berritta2024}, are being developed to address computationally hard problems. In practice, however, any such device operates as an open quantum system coupled to its environment. Modeling this coupling remains a major challenge for spin-based quantum simulation and computation. While the interaction with the environment is necessary for implementing external control and quantum gates, it also induces non-unitary dissipative processes that typically lead to decoherence and loss of quantum correlations. At the same time, the interplay between unitary dynamics and dissipation gives rise to a rich phenomenology, and suitably engineered dissipation can stabilize and prepare non-trivial quantum states \cite{Lidar1998, Beige2000, Kempe2001, Leghtas2013, Budich2015, Roghani2018,liDissipationinducedAntiferromagneticlikeFrustration2021,Wang2023}. A proper theoretical description therefore requires going beyond the wave function formalism to the density matrix representation, whose dynamics, when weakly coupled to a Markovian bath, is governed by the Lindblad master equation.

The exponential growth of the Hilbert space makes exact approaches intractable beyond very small system sizes, creating a pressing need for scalable and efficient approximation methods capable of capturing non-equilibrium dynamics in strongly correlated open quantum systems. Existing numerical methods range from mean-field approaches \cite{Jin2016, Biella2018} to semiclassical \cite{Schachenmayer2015, Huber2021, verstraelenQuantumClassicalCorrelations2023, liMonitoredLongrangeInteracting2025} and tensor network techniques \cite{Vidal2004, verstraete2004renormalizationalgorithmsquantummanybody, Hryniuk2024}. These latter, while highly successful in one dimension (1D), are still significantly less efficient in higher dimensional systems \cite{Kilda2021}. More recently, neural network quantum states have enabled very accurate ground state calculations \cite{Carleo2017, Pfau2024}, yet their extension to non-equilibrium dynamics and steady states remains computationally demanding due to the reliance on Monte Carlo sampling at each time step \cite{hartmannNeuralNetworkApproachDissipative2019,Vicentini2019, Luo2022}. Overall, a common limitation of these approaches is their difficulty in treating large two-dimensional (2D) open spin systems, where both entanglement growth and dissipative dynamics pose severe challenges.

A different route has recently been introduced for bosonic systems with infinite dimensional Hilbert spaces, based on a phase space formulation in which the Wigner function is approximated with a multi-Gaussian ansatz. This approach allows for a fully analytical and sampling-free evaluation of the dynamics \cite{tosca2025VMG}, which exploits automatic differentiation and reaches accuracy comparable to exact diagonalization for much larger system sizes. 
Here, we extend and generalize this phase-space strategy to interacting spin systems by adopting a different representation, namely the Husimi-$Q$ function, and constructing a variational ansatz in terms of a multi-coherent state decomposition. The resulting equations of motion are derived from the Dirac-Frenkel variational principle \cite{Dirac1930} and evaluated analytically, exploiting the phase space structure of the ansatz and automatic differentiation to achieve both high numerical efficiency and precision without Monte Carlo sampling. This approach opens a route to the accurate simulation of non-equilibrium quantum dynamics in large two-dimensional spin lattices.

\begin{figure*}[t]
    \centering
    \includegraphics[width=0.9\textwidth]{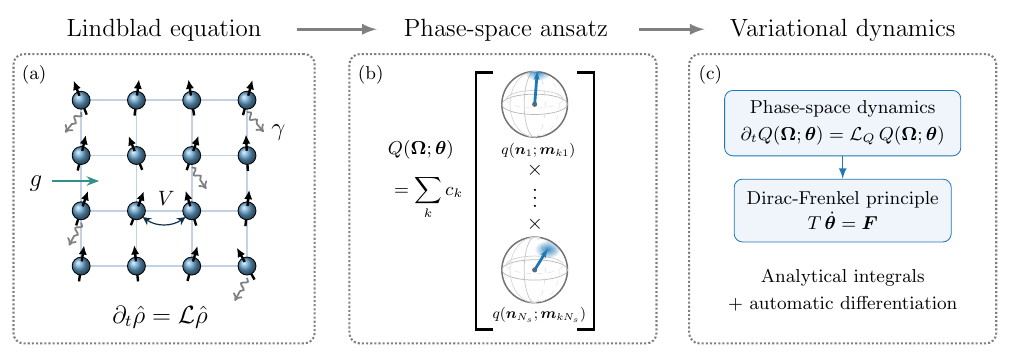}
    \caption{Schematic of the variational phase-space approach introduced in this work.
    (a) A driven-dissipative spin lattice with $N_s$ sites governed by the Lindblad master equation $\partial_t \hat{\rho} = \mathcal{L}\hat{\rho}$, with nearest-neighbour $ZZ$ coupling of strength $V$, transverse-field $g$, and local decay with rate $\gamma$.
    (b) The many-body density matrix is encoded in phase space via the Husimi-$Q$ function, which is parameterized as a variational mixture of $N_c$ spin-coherent-state products: $Q(\bm{\Omega};\bm{\theta}) = \sum_{k=1}^{N_c} c_k \prod_{i=1}^{N_s} q(\bm{n}_i;\bm{m}_{ki})$, where each single-site function $q(\nvec_i;\bm{m}_{ki})$ [cf. Eq.~\eqref{eq:vcms}] lives on the Bloch sphere for site~$i$.
    (c) The phase-space Liouvillian $\mathcal{L}_Q$ generates the dynamics of $Q(\bm{\Omega};\bm{\theta})$; projecting onto the variational manifold via the Dirac-Frenkel principle yields the equations of motion $T\dot{\bm{\theta}} = \bm{F}$, where the quantum geometric tensor $T$ and the force vector $\bm{F}$ are evaluated through analytical integrals [cf. Eq.~\eqref{eq:S-F}] and automatic differentiation, without the need of Monte Carlo sampling.}
    \label{Fig:Scheme}
\end{figure*}

{\it Phase-space representation.---}
We consider a driven-dissipative spin lattice whose dynamics, under weak coupling to a Markovian environment, is governed by the Lindblad master equation~\cite{Breuer2002, Fazio2025}:
\begin{equation}\label{eq:Lindblad}
    \frac {\partial \hat \rho(t)} {\partial t} = \mathcal{L} \hat \rho(t) = - \rmi [\hat H, \hat \rho(t) ] + 
    \sum_j \mathcal{D}_j\hat \rho(t)\,.
\end{equation}
Here, $\mathcal L$ is the Liouvillian superoperator generating the time evolution of the density matrix $\rhohat$. The Hamiltonian $\hat H$ governs the unitary dynamics, while the dissipation superoperators $\mathcal { D}_j$ describe the coupling to the environment through jump operators $\hat \Gamma_j$:
\begin{equation}\label{eq:dissipation}
    \mathcal{ D}_j {\hat \rho(t)}=
 {\hat \Gamma_{j}} {\hat \rho(t)} \hat \Gamma_{j}^{\dagger}-\frac{1}{2}
\left\{
\hat \Gamma_{j}^{\dagger} \hat \Gamma_{j}, 
{\hat \rho(t)}
\right\}\,.
\end{equation}
For a spin-$S$ system, both $\Hhat$ and $\hat\Gamma_j$ are expressed in terms of spin operators $\Shat_i^\alpha$, $\alpha \in \{x,y,z\}$, with $i$ labeling the lattice sites. These satisfy the commutation relations $[\hat S_i^\alpha, \hat S^\beta_j] = \rmi \delta_{ij}\epsilon_{\alpha\beta\lambda} \hat S^{\lambda}$, where $\epsilon_{\alpha \beta \lambda}$ is the Levi-Civita symbol.

The Lindblad dynamics can be equivalently formulated in phase space. This mapping makes use of spin coherent states $\ket{\nvec_i}$ polarized along the unit vector $\nvec_i=(\sin\vartheta_i\cos\varphi_i, \sin\vartheta_i\sin\varphi_i,\cos\vartheta_i)$ on the unit sphere $\mathbb S^2$:
\begin{equation}
    \ket{\nvec_i} = e^{-i \varphi_i \hat S_i^z} e^{-i \vartheta_i \hat S_i^y} \ket{S, m=S}_i,
\end{equation}
where $\ket{S, m=S}_i$ is the eigenstate of $S_i^z$ with maximal projection $m=S$ at site $i$. For a system of $N_s$ spins, the density matrix $\hat \rho$ can be mapped to its Husimi-$Q$ function $Q: (\mathbb S^2)^{N_s} \xrightarrow{} [0,+\infty[$ defined as
 \begin{equation}
    \Omegavec\equiv(\nvec_1,\ldots,\nvec_{N_s}) \mapsto     Q(\Omegavec) = \ncal\bra{\Omegavec} \hat \rho \ket{ \Omegavec}\,,
 \end{equation}
with $\ket{\Omegavec}=\otimes_i\ket{\nvec_i}$ and normalization $\ncal=[(2S+1)/(4\pi)]^{N_s}$ such that $\int Q(\Omegavec)\d\Omegavec =1$.
In this representation, the Lindblad equation~\eqref{eq:Lindblad} becomes a partial differential equation for the Q function, namely \cite{Klimov2002, Zueco2007, KlimovChumakov2009}:
 \begin{equation}\label{eq:phase_space_lindblad}
\begin{aligned}
    \frac{\partial Q(\Omegavec,t)}{\partial t} = \mathcal L_Q Q  =  \{H_Q, Q\}_{MB} + \sum_j {\mathcal{D}_Q}_j[Q],
\end{aligned}
\end{equation}
where $H_Q$ and $\mathcal D_Q$ are the phase-space Hamiltonian and dissipators, and $\{\circ , \circ \}_{MB}$ denotes the Moyal bracket (see End Matter).

{\it Variational ansatz for the Husimi-$Q$ Function.---}
In this work, we adopt a variational ansatz to encode the Husimi-$Q$ function via a mixture of spin coherent states. In what follows, we focus on the case of spin number $S=1/2$; yet, the presented variational approach in phase-space can be generalized to any $S$ and incorporate different variational ans\"atze.

The variational Multi-Coherent State (vMCS) ansatz for a system of $N_s$ spins, based on a mixture of $N_c$ spin-coherent states (see sketch in Fig. \ref{Fig:Scheme}), reads:
\bea\label{eq:vcms}
    Q(\Omegavec;\thetavec) &= \sum_{k=1}^{N_c} c_k\prod_{i = 1}^{N_s} q(\nvec_i;\mvec_{ki})\,,\\
    q(\nvec_i;\mvec_{ki})&= \frac{1}{4 \pi} (1 + {\nvec}_i \cdot {\mvec}_{ki})\,,
\eea
where the variational parameters are $\thetavec \equiv (\{c_k\}, \{\boldsymbol m_{ki}\})$. 
Each component (indexed by $k$) is fully specified by $\mvec_{ki}$. When $\lVert \mvec_{ki}\rVert = 1$, it represents a spin coherent state polarized along the three-dimensional vector $\mvec_{ki}$. We also allow $\lVert \mvec_{ki}\rVert\leq 1$ to include spin thermal states~\footnote{For spin-$1/2$, they are mixtures of two spin coherent states with opposite polarizations.} within the ansatz.
The normalization of the Husimi-$Q$ function imposes the constraint
\begin{equation}
    \sum_{k=1}^{N_c}c_k = 1 \, {\rm with} \, \, c_k \in \mathbb{R}.
\end{equation}
{\it Importantly, the coefficients $c_k$ can be negative}, which is essential for capturing quantum correlations. This feature fundamentally distinguishes the v-MCS approach from trajectory-based semiclassical methods, which by construction are restricted to \textit{positive} mixtures of semiclassical states.

The dynamical evolution of the parameters $\thetavec$ is obtained from the Dirac-Frenkel variational principle \cite{Lasser2022}. 
This amounts to minimizing, with respect to ${\dot \thetavec}$, the cost function $\mathcal J({\dot \thetavec})$, defined as
\begin{equation}
    \mathcal J(\dot \thetavec) = \lVert\mathcal L_Q Q - \partial_t Q \rVert_2^2 = \lVert\mathcal L_Q Q - \partial_\thetavec Q^T \dot \thetavec \rVert_2^2, 
\end{equation}
where $\lVert \cdot\rVert_2$ denotes the $L^2$ norm, and $\partial_t Q$ has been projected onto the tangent space of the variational manifold. We thus obtain the equations of motion for the variational parameters:
\begin{equation}
     T~{\dot \thetavec} = \Fvec,
\end{equation}
where 
\bea\label{eq:S-F}
T_{kl} &= \int_{(\mathbb{S}^2)^{N_s}} \frac{\partial Q}{\partial{\theta_k}} \frac{\partial Q}{\partial{\theta_l}} \d\Omegavec\,,\\
    F_k &= \int_{(\mathbb{S}^2)^{N_s}} \frac{\partial Q}{\partial \theta_k}  \mathcal L_Q Q ~\d\Omegavec\,. 
\eea
Importantly, both the quantum geometric tensor $T$ and the force vector $\Fvec$ can be computed analytically for any mixture of spin coherent states (see End Matter). 
This eliminates the need for sampling-based estimations of $T$ and $\Fvec$, enabling smooth and high-precision simulations of the dynamics \footnote{In practice, for numerical stability, we employ Tikhonov regularization when inverting the matrix $T$. This amounts to adding a small diagonal shift $\epsilon$ (typically in the range $[10^{-11},10^{-9}]$), leading to the modified equation ($T + \epsilon \mathbbm 1)\dot \thetavec = \boldsymbol F$.}.

The ansatz can be further enhanced by incorporating prior knowledge from the physical system, such as the presence of translational symmetry. Denoting by $\Gcal$ the group of lattice-site permutations that leave the Liouvillian invariant, we define a symmetrized v-MCS ansatz,
\bea\label{eq:v-CMS-symm}
    Q_\Gcal(\Omegavec;\thetavec) &= \frac{1}{|\Gcal|}\sum_{\tau\in\Gcal}\sum_{k=1}^{N_c}c_k\prod_{i=1}^{N_s}q(\nvec_i;\mvec_{k\tau(i)})\,,
\eea
which efficiently captures the dynamics and the steady state sharing the symmetries of the system.

{\it Results.---}
To assess the expressive power of the v-MCS  ansatz, we study the driven-dissipative quantum dynamics and the resulting non-equilibrium steady state of the one- and two-dimensional dissipative transverse-field Ising model. This problem has become a standard benchmark for neural network approaches \cite{Vicentini2019, Luo2022, Mellak2024}, enabling direct comparison with sampling-based methods. The corresponding Hamiltonian reads 
\begin{equation}
    \hat H = \sum_i \frac{g}{2} \hat \sigma_i^x + \frac{V}{2 \chi} \sum_{\langle i,j \rangle} \hat \sigma_i^z \hat \sigma_{j}^z\,,
\end{equation}
where $\sigma_i^\alpha$, with $\alpha \in \{x,y,z\}$, are Pauli matrices acting on site $i$, related to spin-$1/2$ operators via $\Shat^\alpha_i=\sigmahat^\alpha_i/2$. 
The parameter $V$ denotes the nearest-neighbor $ZZ$ coupling strength, while $g$ is the strength of the transverse magnetic field along the $x$ direction. The factor $\chi$ is the lattice coordination number.
Each spin undergoes a local decay process described by the jump operator $\hat \Gamma_i = \sqrt{\gamma}\hat \sigma_i^- = \frac{\sqrt{\gamma}}{2}(\hat \sigma_i^x - \mathrm i \hat \sigma_i^y)$, where $i$ labels the lattice site and $\gamma$ is the decay rate. 
In the following, we consider the model on a $L_x\times L_y$ lattice with periodic boundary conditions, and incorporate these symmetries into the symmetrized v-MCS ansatz~\eqref{eq:v-CMS-symm}.

\begin{figure}[t]
    \centering
    \includegraphics[width=\columnwidth]{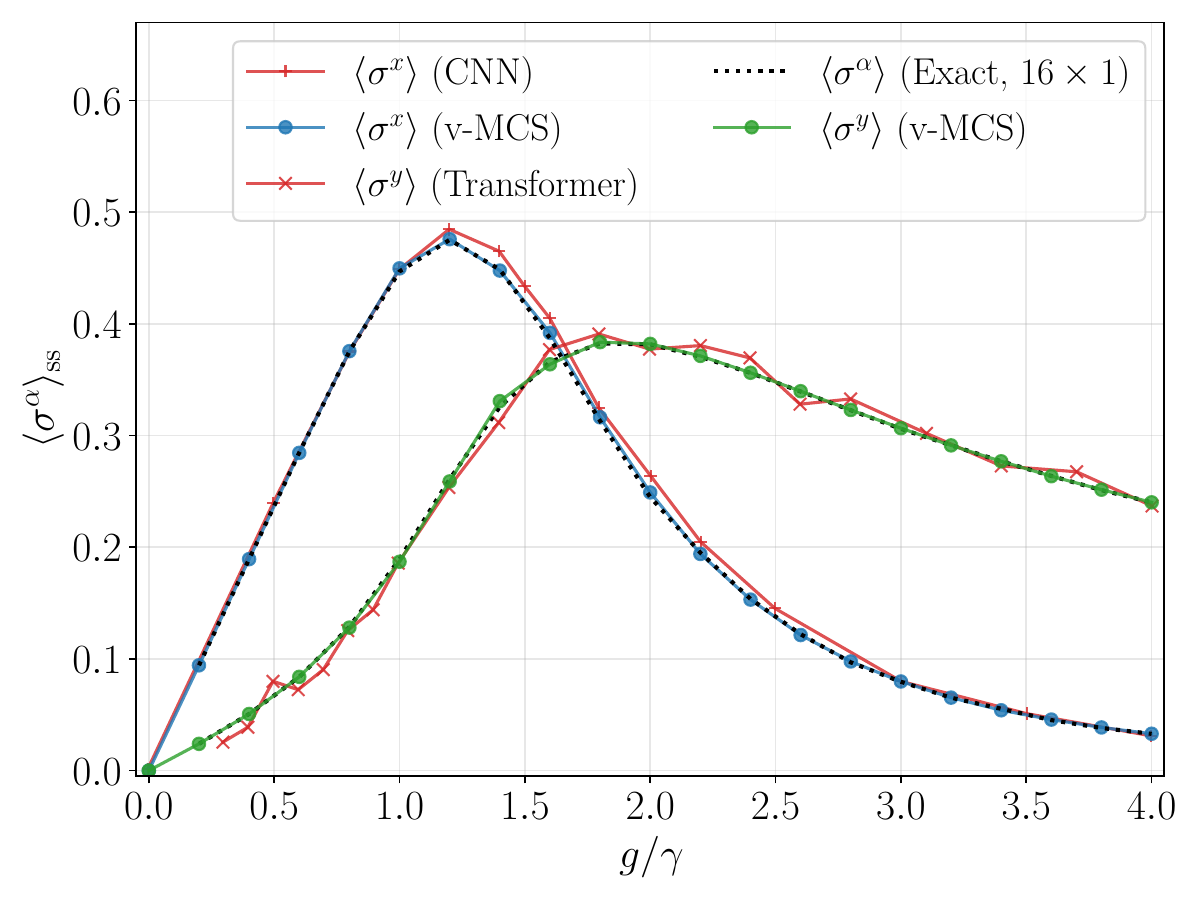}
    \caption{Variational steady-state results for the observables $\langle \hat \sigma_x\rangle$ and $\langle \hat \sigma_y\rangle$ for a $1$D $16\times 1$ dissipative transverse-field Ising model. The parameter $g$ represents the external magnetic field along $X$ while $\gamma$ is the loss rate associated to the jump operator $\hat \Gamma_i = \sqrt{\gamma} \hat \sigma_i^-$. The v-MCS is compared to neural-network steady-state results with Transformer and Convolutional Neural Networks (CNN) architectures (data taken respectively from \cite{Luo2022} and \cite{Mellak2024}, see legend). Exact results are obtained via Monte Carlo wave function simulations. The v-MCS ansatz at convergence has $N_c =16$ total components per spin for a total of $784$ parameters. The initial condition for the dynamics is  $\prod_{i=1}^{N_s}\ket{\uparrow}_x$. The $ZZ$ coupling strength is $V /\gamma = 2$.}
    \label{fig:T_vs_v_MCS}
\end{figure}
We study both one- and two-dimensional systems. We first benchmark the steady-state results of the variational dynamics against the numerically exact solution for a $16 \times 1$ spin lattice. This is performed over multiple values of the transverse-field $g/\gamma$, with fixed coupling $V/\gamma = 2$. We employ a v-MCS ansatz with $N_c=16$ components per spin, corresponding to a total of $784$ variational parameters ($3 \times N_c \times N_s + N_c$). Expectation values of observables are computed via their $P$-symbols in phase space (see End Matter).
In Fig.~\ref{fig:T_vs_v_MCS}, we report the v-MCS steady-state results for $\langle \hat \sigma^x\rangle$ and $\langle \hat \sigma^y\rangle$ (we drop the site index $i$ due to the translational invariance). Remarkably, the variational predictions are in perfect agreement with the exact solution obtained via Monte Carlo wave function simulations. We also compare with results from state-of-the-art neural network approaches, namely Convolutional Neural Networks (CNN) \cite{Mellak2024} and Transformers \cite{Luo2022}. Fig.\ref{fig:T_vs_v_MCS} shows that these methods do not achieve the same level of accuracy for the non-equilibrium steady state as the v-MCS approach.
This performance can be attributed to the absence of sampling in the v-MCS method, which yields smooth dynamics and efficient numerical evaluation. In practice, a typical simulation for the $1$D $16$-spin transverse-field Ising model, for a given value of $g$, requires about $1$ minute on a standard desktop computer (Mac mini with Apple M4 Pro processor), computing the full quantum evolution up to $t_f = 31\,\gamma^{-1}$, from which the steady state is extracted.

\begin{figure}[t]
    \centering
    \includegraphics[width=\columnwidth]{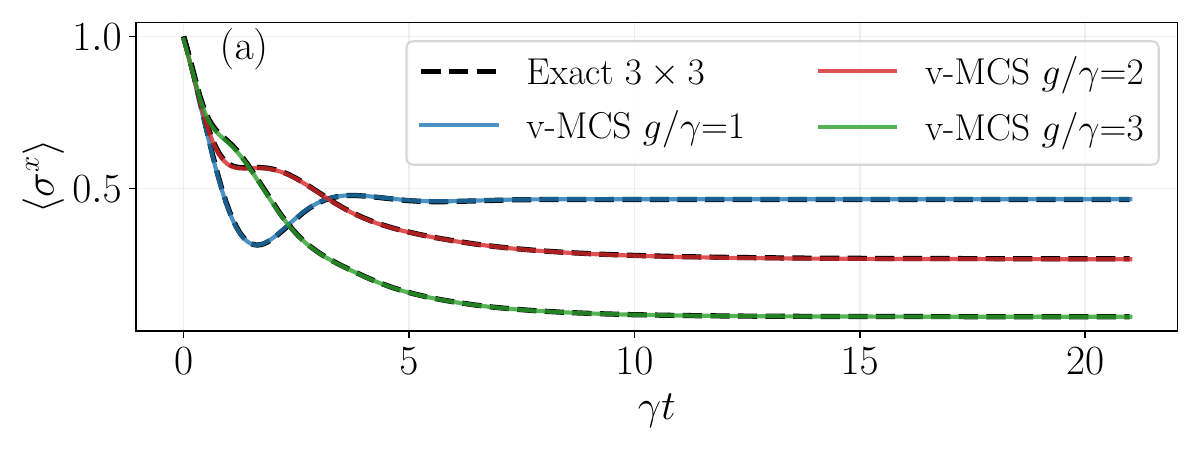}
    
    \vspace{-0.2cm}
    
    \includegraphics[width=\columnwidth]{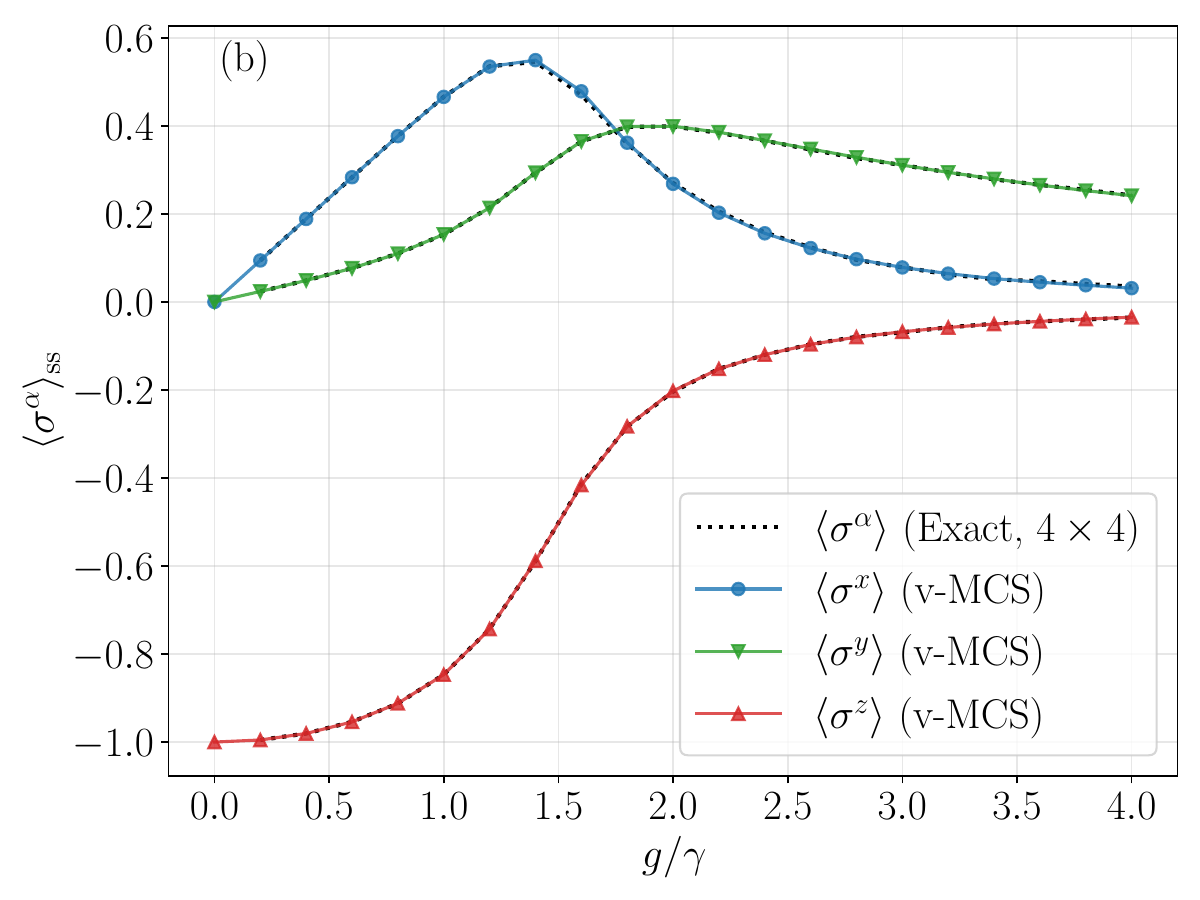}
    
    \caption{Benchmark results in 2D spin lattices. (a) Variational real-time dynamics (solid lines) of a $3\times 3$ dissipative transverse-field Ising model for $g/\gamma = 1, 2, 3$, compared with the exact solution (dashed lines). The v-MCS ansatz uses $N_c = 10$ components per spin ($280$ variational parameters). The initial state is a small perturbation of $\prod_{i=1}^{N_s}\ket{\uparrow}_x$.
    (b) Steady-state expectation values $\langle \hat\sigma^\alpha \rangle_\mathrm{ss}$ for a $4\times 4$ lattice as a function of $g/\gamma$. The v-MCS results ($N_c = 16$, $784$ parameters) are compared with exact results (see legend). In both panels, $V/\gamma = 2$.}
    \label{fig:2d_steady}
\end{figure}

The v-MCS approach, moreover, immediately generalizes, without any modification, to any lattice connectivity, being only limited by its expressive power, namely, the total number of variational parameters. 
To demonstrate this, we first benchmark the full real-time dynamics of 2D transverse-field Ising model obtained with the v-MCS method. In Fig.~\ref{fig:2d_steady} (a), we show the results on a $3\times 3$ lattice for three values of the transverse-field, $g/\gamma = 1, 2, 3$. The variational time evolution perfectly reproduces both the transient dynamics and the steady state. These results are obtained with $N_c = 10$ components per spin ($280$ variational parameters), and the full variational quantum dynamics up to $\gamma t_f = 21$ is computed in approximately $10$ seconds on a personal computer. This is in sharp contrast with neural-network-based methods, which struggle in accurately capturing real-time dynamics in two-dimensional spin lattices~\cite{Luo2022, hartmannNeuralNetworkApproachDissipative2019}. 
we further benchmark the steady-state expectation values for a $4\times4$ lattice over multiple values of $g/\gamma$, with $V/\gamma = 2$.
As shown in Fig.~\ref{fig:2d_steady} (b), the ansatz perfectly captures the steady-state values with $N_c = 16$ components per spin ($768$ variational parameters, as in the $1$D case).

\begin{figure}[t]
    \centering        
    \includegraphics[width=\columnwidth]{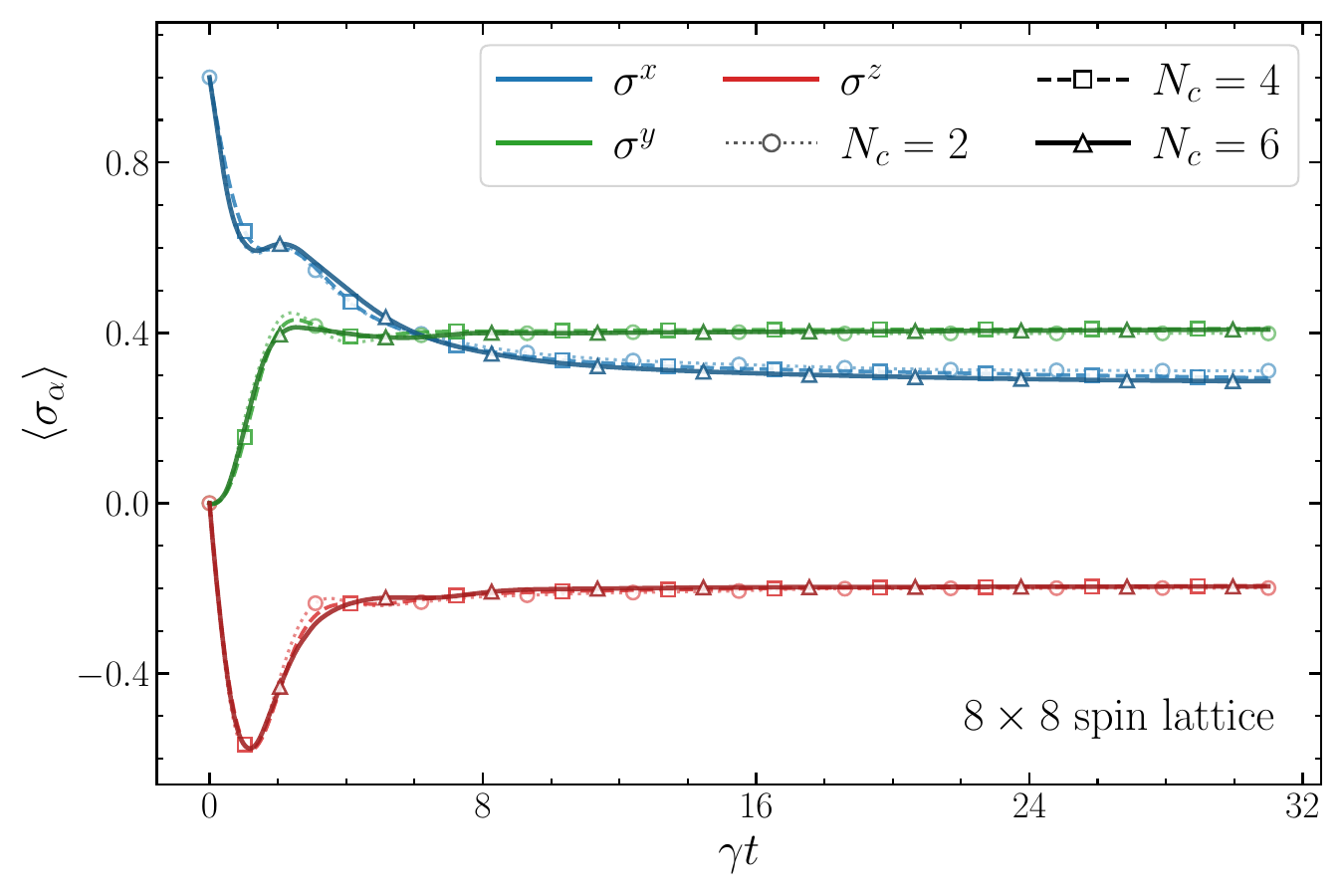}
    \caption{Variational dynamics of $\langle\hat\sigma^\alpha\rangle$, with $\alpha \in \{x,y,z\}$ for the $8\times 8$ dissipative transverse-field Ising model, computed with $N_c \in \{2, 4, 6\}$ coherent-state components (respectively with $386,772$ and $1158$ variational parameters). System parameters: $V/\gamma = 2$, $g/\gamma = 2$. The initial state is $\prod_{i=1}^{N_s}\ket{\uparrow}_x$. Both transient dynamics and steady-state values exhibit systematically controlled convergence as $N_c$ increases.}
\label{fig:dynamics}
\end{figure}

Finally, in Fig.~\ref{fig:dynamics}, we present the variational dynamics for a $2$D $8\times 8$ lattice ($64$ spins) with $g/\gamma = 2$ and $V/\gamma = 2$. The ansatz exhibits controlled convergence of both transient and steady-state observables as the number of components $N_c$ increases. Notably, the steady state is already accurately captured with $N_c = 2$, corresponding to only $386$ variational parameters. Simulations of the $2$D $8\times 8$ lattice were performed on H100 GPU with a typical simulation-time of $10$ minutes. 

{\it Conclusion and Perspectives.---}
We have introduced an efficient variational method to simulate the driven-dissipative dynamics of interacting spin systems, based on a phase-space representation of the many-spin density matrix via the Husimi-$Q$ function. By parameterizing it with a mixture of spin-coherent states (v-MCS ansatz), whose coefficients are allowed to take negative values, the method is able to faithfully capture genuine quantum correlations beyond semiclassical descriptions. Combined with the fully analytical evaluation of the variational equations of motion, this approach bypasses stochastic sampling while achieving controlled convergence of both transient dynamics and non-equilibrium steady states.

We benchmarked the approach on the dissipative transverse-field Ising model in one and two dimensions, finding excellent agreement with exact results and improved accuracy over state-of-the-art neural-network architectures, at a fraction of the computational cost. The method further extends straightforwardly to arbitrary lattice geometries and dimensionalities.

The v-MCS ansatz establishes a sampling-free paradigm for variational simulations of dissipative spin dynamics, opening several promising directions. Its extension to higher-spin systems ($S > 1/2$) is conceptually straightforward within the same phase-space framework, requiring adapted ansätze on generalized Bloch spheres. 
Moreover, the ability to efficiently capture real-time dynamics makes it well-suited to exploring intrinsically dynamical phenomena in open quantum systems, such as dissipative time crystals~\cite{ieminiBoundaryTimeCrystals2018, Cabot2022, delmonteMeasurementinducedPhaseTransitions2025}. Finally, the analytical structure of the framework can be naturally combined with different variational ansatz families—for instance, by incorporating inter-spin correlations directly into the ansatz components—to efficiently parameterize strongly entangled regimes and critical points of dissipative phase transitions. 

We acknowledge support from the French ANR project FracTrans (grant ANR-24-CE30-6983) and a grant (Polaritonic) from the French Government managed by the ANR under the France 2030 programme with the reference ANR-24-RRII-0001. This work was granted access to the HPC resources of IDRIS under the allocation 2025-AD010612462R2 and A0190916893 made by GENCI.

\bibliography{biblio}

\onecolumngrid

\section{End Matter}
\label{EndMatter}
\setcounter{equation}{0}
\setcounter{figure}{0}
\setcounter{table}{0}
\makeatletter
\renewcommand{\theequation}{A\arabic{equation}}
\renewcommand{\thefigure}{A\arabic{figure}}
\twocolumngrid

{\it Variational phase-space equations of motion.---}
We summarize here the phase-space formulation of spin dynamics for a single spin-$S$. The generalization to $N_s$ spins follows directly from the Cartesian-product structure of the phase space $\mathbb{S}^2\times\cdots\times \mathbb{S}^2$.

The action of the operator $\hat S^z$ is represented in phase space by the operator acting on the Husimi-$Q$ function~\cite{KlimovChumakov2009}:
\begin{equation}
    \left.\begin{array}{l} \hat \rho \hat S^z \\ \hat S^z \rho \end{array}\right\} \leftrightarrow [\mp \frac{l_z}{2} + \Lambda_0(\varphi, \vartheta) ]Q(\varphi, \vartheta)\,,
\end{equation}
where $(\varphi, \vartheta)$ are spherical coordinates, and
\begin{equation}
    l_z = -i \partial_\varphi\,,~
    \Lambda_0 = \frac{1}{2}\left[ 2S \cos \vartheta - \sin \vartheta \partial_\vartheta \right]\,.
\end{equation} 
The ladder operators $\hat S^\pm$ admit analogous phase-space representations~\cite{KlimovChumakov2009}:
\begin{equation}
    \left.\begin{array}{l} \hat \rho \hat S^+ \\ \hat S^+ \hat \rho \end{array}\right\} \leftrightarrow [\mp l_{+} / 2+\Lambda_{+}(\varphi, \vartheta)]Q(\varphi,\vartheta)\,, 
\end{equation}
\begin{equation}
    \left.\begin{array}{l} \hat \rho \hat S^- \\ \hat S^- \hat \rho \end{array}\right\} \leftrightarrow [\mp l_{-} / 2+\Lambda_{-}(\varphi, \vartheta)]Q(\varphi,\vartheta)\,,
\end{equation}
with
\begin{equation}
    \Lambda_{ \pm}=\mathrm{e}^{ \pm i \varphi} \frac{\sin \vartheta}{2 \varepsilon} \pm \frac{1}{2}\left[\cos \vartheta l_{ \pm}-\mathrm{e}^{ \pm i \varphi} \sin \vartheta\left(l_z \pm 1\right)\right]\,,
\end{equation}
and 
\begin{equation}
    l_{ \pm}=\mathrm{e}^{ \pm i \varphi}\left( \pm \partial_\vartheta+i \cot \vartheta \partial_\varphi\right)\,,
\end{equation}
where $\varepsilon=(2S+1)^{-1}$.
As a simple visualization of the phase-space representation of the dynamics, let us compute the coherent spin dynamics generated by the Hamiltonian $\hat H =g \hat S^x = {g}(\hat S^+ +\hat S^-)/2$.
The action of Hamiltonian $-\rmi[\Hhat,\bullet]$ then translates into the Moyal bracket $\{ H_Q,\bullet \}_{MB}$, using the above mapping rules:

\bea
    \{H_Q,Q\}_{MB} &= -\rmi \ g/2\big( [+ l_{+} / 2+\Lambda_{+}(\varphi, \vartheta)]Q\\&\phantom{===} + [l_{-} / 2+\Lambda_{-}(\varphi, \vartheta)]Q\\
    &\phantom{===}- [- l_{+} / 2+\Lambda_{+}(\varphi, \vartheta)]Q\\
    &\phantom{===}- [- l_{-} / 2+\Lambda_{-}(\varphi, \vartheta)]Q\big)\\
    &=g(\sin \varphi \partial_{\vartheta}+\cot \vartheta  \cos \varphi \partial_{\varphi} )Q(\vartheta ,\varphi )\,,
\eea
which effectively rotates the $Q(\vartheta,\varphi)$ function around the $x$ axis, as expected.

The computation of the quantum geometric tensor $T_{kl}$ and the force vector $F_k$ requires evaluating the integrals defined in Eq.~\eqref{eq:S-F}. These can be recast in the equivalent form:
\bea
    T_{kl}(\thetavec) &= \partial_{\theta^L_k}\partial_{\theta^R_l}\Ical_T(\thetavec^L,\thetavec^R)|_{\thetavec^{L}=\thetavec^R=\thetavec}\,,\\
    F_k(\thetavec) &= \partial_{\theta^L_k}\Ical_F(\thetavec^L,\thetavec^R)|_{\thetavec^{L}=\thetavec^R=\thetavec}\,,
\eea
where the overlap integrals are defined as
\bea
    \Ical_T(\thetavec^L,\thetavec^R) &\equiv \int_{(\mathbb{S}^2)^{N_s}}Q(\Omegavec;\thetavec^L)Q(\Omegavec;\thetavec^R)\d\Omegavec\\
    &= \langle Q(\thetavec^L),Q(\thetavec^R)\rangle\,,\\
    \Ical_F(\thetavec^L,\thetavec^R) &\equiv \int_{(\mathbb{S}^2)^{N_s}} Q(\Omegavec;\thetavec^L)\lcal_Q Q(\Omegavec;\thetavec^R)\d\Omegavec\\
    &= \langle Q(\thetavec^L),\lcal_Q Q(\thetavec^R)\rangle\,,
\eea
with the shorthand notation
\bea
    Q(\thetavec) \equiv ( \Omegavec \mapsto Q(\Omegavec;\thetavec) )
\eea
and the $L^2$ inner product defined as
\bea
    \langle f,g \rangle \equiv \int_{(\mathbb{S}^2)^{N_s}} f(\Omegavec)g(\Omegavec)\d\Omegavec\,.
\eea
Here, $\thetavec^{L,R}$ are auxiliary variables that allow derivatives to be taken outside the generating function integrals $\Ical_{T,F}$, and are set equal at the end. Importantly, all integrals above can be evaluated \textit{analytically}, completely avoiding the stochastic sampling required in other variational approaches.

We now derive the equations of motion for the model considered in the main text within the v-MCS ansatz for the Husimi-$Q$ function, focusing on spin $S=1/2$. 
Given the structure of the ansatz~\eqref{eq:vcms}, 
\bea
    Q(\thetavec) &= \sum_{k=1}^{N_c} c_k\prod_{i=1}^{N_s} q(\mvec_{ki})\,,
    \eea
the variational parameters are $\thetavec=\{c_k,\mvec_{ki}\}$, and the integrals introduced above simplify accordingly. The first integral evaluates to
\bea
    \Ical_T(\thetavec^L,\thetavec^R) &= \sum_{kl}c^L_k c^R_l \prod_i\langle q(\mvec^L_{ki}),q(\mvec^R_{li})\rangle\,,
\eea
where the local overlap integral on a single Bloch sphere is
\bea
    \langle q(\mvec^L),q(\mvec^R)\rangle &= \int_{\mathbb{S}^2}q(\nvec;\mvec^L)q(\nvec;
    \mvec^R)\d\mu(\nvec)\\&= \dfrac{1}{12\pi}(3+\mvec^L\cdot\mvec^R)\,.
\eea
The second integral is model dependent, yet remains tractable due to the local structure of the Liouvillian. In particular, $\lcal$ can be decomposed into one- and two-local contributions:
\bea
    \lcal = \sum_{i}\lcal_i^{(1)} + \sum_{\langle i,j\rangle}\lcal_{ij}^{(2)}\,,
\eea
with
\bea
    \lcal_i^{(1)} &= -\rmi g[\Shat_i^x,\bullet] + \gamma \left(\Shat_i^-\bullet\Shat_i^+ - \dfrac{1}{2}\left\{ \Shat_i^+\Shat_i^-,\bullet \right\}\right)\,,\\
    \lcal_{ij}^{(2)} &= -\rmi V[\Shat_i^z\Shat_j^z,\bullet]\,.
\eea
It follows that the relevant contributions reduce to the evaluation of the following integrals:
\bea
    &\langle q(\mvec_i^L), \lcal^{(1)}_{i,Q} q(\mvec_i^R)\rangle \\&= \dfrac{g}{12\pi}(\mvec_i^R\times\mvec_i^L)_x\\
    &-\dfrac{\gamma}{24\pi}[\mvec_i^L\cdot\mvec_i^R + m^L_{i,z}(2+m^R_{i,z})]\,,
\eea
\bea
     &\langle q(\mvec_i^L)q(\mvec_j^L), \lcal^{(2)}_{ij,Q} q(\mvec_i^R)q(\mvec_j^R)\rangle \\&= \dfrac{V}{288 \pi ^2} \left[(m^L_{j,z}+3 m^R_{j,z}) (\mvec^R_i\times\mvec^L_i)_z\right.\\
     &\phantom{=====}\left.+(m^L_{i,z}+3 m^R_{i,z}) (\mvec^R_j\times\mvec^L_j)_z\right]\,.
\eea
These expressions lead to the Liouvillian overlap integral generating the force vector:
\bea
    &\Ical_F(\thetavec^L,\thetavec^R) = \sum_{kl}c_k^L c_l^R \prod_\mu\langle q(\mvec^L_{k\mu}),q(\mvec^R_{l\mu})\rangle\\
    &\times\left\{\sum_i\dfrac{\langle q(\mvec^L_{ki}),\lcal^{(1)}_{i,Q}q(\mvec^R_{li}) \rangle}{\langle q(\mvec^L_{ki}),q(\mvec^R_{li}) \rangle}\right.\\
    &\left.+\sum_{\langle i,j\rangle} \dfrac{\langle q(\mvec^L_{ki})q(\mvec^L_{ij}),\lcal^{(2)}_{ij,Q}q(\mvec^R_{li})q(\mvec^R_{lj}) \rangle}{\langle q(\mvec^L_{ki}),q(\mvec^R_{li}) \rangle \langle q(\mvec^L_{kj}),q(\mvec^R_{lj}) \rangle}\right\}\,.
\eea
The quantum geometric tensor $T_{kl}$ and the force vector $F_k$ are then obtained by analytically differentiating this expression with respect to the variational parameters, which can be performed efficiently, for instance, using automatic differentiation.

\setcounter{equation}{0}
\setcounter{figure}{0}
\setcounter{table}{0}
\makeatletter
\renewcommand{\theequation}{B\arabic{equation}}
\renewcommand{\thefigure}{B\arabic{figure}}
\twocolumngrid

{\it Phase-space evaluation of observables.---}
In the phase-space formulation of quantum mechanics, the expectation value of an observable $\Ahat$ can be obtained via its $P-$symbol, which is a phase-space function $P_A(\Omegavec)$ defined as~\cite{KlimovChumakov2009}
\bea
    \Ahat = \left(\dfrac{2S+1}{4\pi}\right)^{N_s}\int\ketbra{\Omegavec} P_{\Ahat}(\Omegavec) \d \Omegavec\,,
\eea
such that one has,
\bea
    \Tr[\rhohat\Ahat]&=\left(\dfrac{2S+1}{4\pi}\right)^{N_s}\int \bra{\Omegavec}\rhohat\ket{\Omegavec} P_{\Ahat}(\Omegavec) \d \Omegavec\\
    &= \int Q(\Omegavec) P_{\Ahat}(\Omegavec)\d\Omegavec\,,
\eea
where the integrals above are performed over ${(\mathbb{S}^2)^{N_s}}$. In particular, we have
\bea
    P_{\idhat}(\Omegavec) = 1\,,~
    P_{\Shat_i^\alpha}(\Omegavec) = (S+1)n_i^\alpha\,,
\eea
and that the $P$ symbol of the product of commuting operators (such as those acting on different spins) is simply the product of the individual $P$ symbols. Therefore, for our v-MCS ansatz (with $S=1/2$), the expectation value of a single-site spin operator is simply,
\bea
    \langle\Shat_i^\alpha\rangle &= \int Q(\Omegavec) P_{\Shat^\alpha_i}(\Omegavec) \d\Omegavec\\
    &= \sum_{k=1}^{N_c} c_k\int_{\mathbb{S}^2} \dfrac{1}{4\pi}(1+\nvec\cdot\mvec_{ki})\left(\frac{1}{2}+1\right)n^\alpha_i \d \Omega(\nvec)\\
    &= \dfrac{1}{2}\sum_kc_k m^\alpha_{ki}\,,
\eea
where we recall the notations defined in the main text that $\Omegavec\equiv(\nvec_1,\ldots,\nvec_{N_s})$, and $\d\Omegavec$ denotes the spherical integration measure.

\end{document}